\documentclass{article}
\usepackage{spconf,amsmath,graphicx,booktabs,hyperref}
\usepackage{cite}

\title{GTN-Bailando: Genre Consistent Long-Term 3D Dance Generation based on Pre-trained Genre Token Network}
\name{
\begin{tabular}{@{}c@{}}
Haolin Zhuang$^{1,*}$\thanks{* Work conducted when the first author was intern at XVerse Inc.}, Shun Lei$^{1}$, Long Xiao$^{1}$, Weiqin Li$^{1}$, Liyang Chen$^{1}$,\\
\textit{Sicheng Yang$^{1}$, Zhiyong Wu$^{1,3,\dagger}$\thanks{$\dagger$ Corresponding author.}, Shiyin Kang$^{2,\dagger}$, Helen Meng$^{3}$}
\end{tabular}
}
\address{  $^1$Shenzhen International Graduate School, Tsinghua University, Shenzhen, China\\
  $^2$XVerse Inc., Shenzhen, China\\
  $^3$The Chinese University of Hong Kong, Hong Kong SAR, China}
  
 \address{
    $^1$ Shenzhen International Graduate School, Tsinghua University, Shenzhen, China\\
    $^2$ XVerse Inc., Shenzhen, China\\
    $^3$ The Chinese University of Hong Kong, Hong Kong SAR, China\\
    \small{
        \{zhuanghl21, leis21\}@mails.tsinghua.edu.cn, 
        zywu@sz.tsinghua.edu.cn,
        kangshiyin@xverse.cn,
    }
}

\begin{document}
\ninept
\maketitle
\begin{abstract}
Music-driven 3D dance generation has become an intensive research topic in recent years with great potential for real-world applications. Most existing methods lack the consideration of genre, which results in genre inconsistency in the generated dance movements. In addition, the correlation between the dance genre and the music has not been investigated. To address these issues, we propose a genre-consistent dance generation framework, GTN-Bailando. 
First, we propose the Genre Token Network (GTN), which infers the genre from music to enhance the genre consistency of long-term dance generation. Second, to improve the generalization capability of the model, the strategy of pre-training and fine-tuning is adopted. 
Experimental results on the AIST++ dataset show that the proposed dance generation framework outperforms state-of-the-art methods in terms of motion quality and genre consistency\footnote{Generated demo: \href{https://im1eon.github.io/ICASSP23-GTNB-DG/}{https://im1eon.github.io/ICASSP23-GTNB-DG/}}.

\end{abstract}
\section{Introduction}
\label{sec:intro}

Dance genres~\cite{aist-dance-db} are the manifestations that arise from dance, each with its specific characteristics.
Diversified dance genres increase the variety of dance, making it one of the most well-known forms of artistic expression globally.
With the advancement of artificial intelligence~(AI), it is possible to generate dance pose sequences for promising applications such as choreography assistance and virtual idol performance.
However, producing a satisfactory dance pose sequence using AI choreographies is still challenging due to the difficulty of maintaining genre consistency.
In a certain dance genre, not all physically possible dance poses are appropriate,
and the choreographed dance poses have stricter positional restrictions
as well as a correlation to the given music.

Most existing AI choreographies focus on the alignment between dances and the musical melody.
Early methods~\cite{alemi2017groovenet, TaoranTang2018DanceWM, NelsonYalta2019WeaklySupervisedDR, HsuanKaiKao2020TemporallyGM, ahn2020generative, li2020learning, RuoziHuang2020DanceRL, li2021dance, valle2021transflower, zhang2022music} send music and dance directly into a single network to generate dance sequences autoregressively.
However, such methods suffer from the accumulation of errors during autoregressive generation and will likely regress non-dancing movements.
Meanwhile, some works~\cite{fan2011example, lee2013music, lee2019dancing, KangChen2021ChoreoMasterCM, ZijieYe2020ChoreoNetTM, ofli2011learn2dance} use the retrieval method to divide dances into fixed-length units and choreograph by splicing these units according to the melody of the music.
Although these methods ensure the quality of the generated dance, they are incompatible with different time signatures and beats per minute (BPM).
Recently, a great dance generation framework, Bailando~\cite{siyao2022bailando}, uses the vector quantised-variational autoencoder~(VQ-VAE)~\cite{AaronvandenOord2017NeuralDR} with the Actor-Critic~\cite{VijayRKonda2002ActorcriticA} generative pre-trained transformer~(GPT)~\cite{radford2019language} 
to solve problems in both autoregressive and retrieval dance generation methods.




However, all the studies mentioned above ignore the genre during the dance generation.
Without genre, the generated dance will perform multiple genres of dance movements in a single music clip, resulting in performances that are inappropriate with the music (e.g., ballet movements in hip-hop music).
Thereby, several latest dance generation frameworks devote efforts to the genre.
GCDG~\cite{9747838} uses a one-hot vector to represent the genre and embeds the genre into the transformer decoder to generate genre consistency dance. 
MNET~\cite{kim2022brand} uses a mapping network to transform a latent code into the style code for multi-genres while generating dance motions.
These approaches can generate dance with a particular genre but require auxiliary inputs in inference, such as manually-determined genre labels.
In addition, there is a specific correlation between the dance genre and its background music, 
allowing choreographers to determine the genre of dance based on the music.
Therefore, it is preferable for AI choreographies to infer the genre based on the music and generate dances using the genre that has been inferred.

In this paper, we propose a genre consistent long-term dance generation framework based on Genre Token Network (GTN).
First, we introduce GTN, to infer the genre by learning the correlation between music and genres. 
For a given clip of music, the genre can be inferred through GTN and used as a condition in dance generation, ensuring each generated dance pose satisfies the positional restrictions of one genre.
Second, we propose a strategy for pre-training GTN to improve its generalization capacity. 
Due to the insufficiency of music data in existing dance-music-aligned datasets, it is difficult for the GTN to infer genre from music accurately.
Thereby, in order to strengthen the correlation between each genre and its corresponding music, we collect a large-scale of dance background music with genre labels to pre-train the GTN.
After that, we use a dance-music-aligned dataset to train the dance generation framework and load the pre-trained GTN weights for fine-tuning.
Thus, the GTN can infer the genre of the music more effectively, which further enhances genre consistency.
Experimental results show that our proposed dance generation framework and pre-training strategy significantly outperforms in both evaluation metrics and visualization judgments. 


\begin{figure*}[th]
\centering
\centerline{\includegraphics[width=17cm]{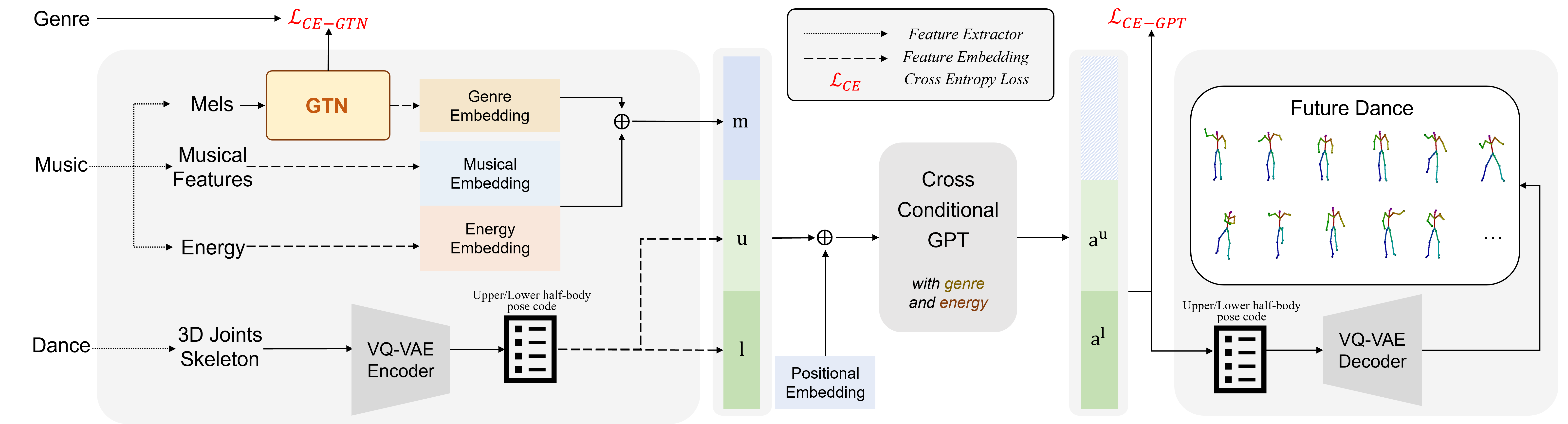}}
\caption{GTN-Bailando. Mel-spectrogram of music is fed to Genre Token Network (GTN). 
The generated genre embedding is utilized as a condition in GPT to guarantee that each generated dance is consistent with its genre.
The parameters are learned via cross-entropy loss $\mathcal{L}_{C E-GTN}$ with genre label and $\mathcal{L}_{C E-GPT}$ with ground-truth dance pose code.}
\label{fig:framework}
\end{figure*}

\vspace{-0.1cm}

\begin{figure}[t]
\centering
\centerline{\includegraphics[width=8.5cm]{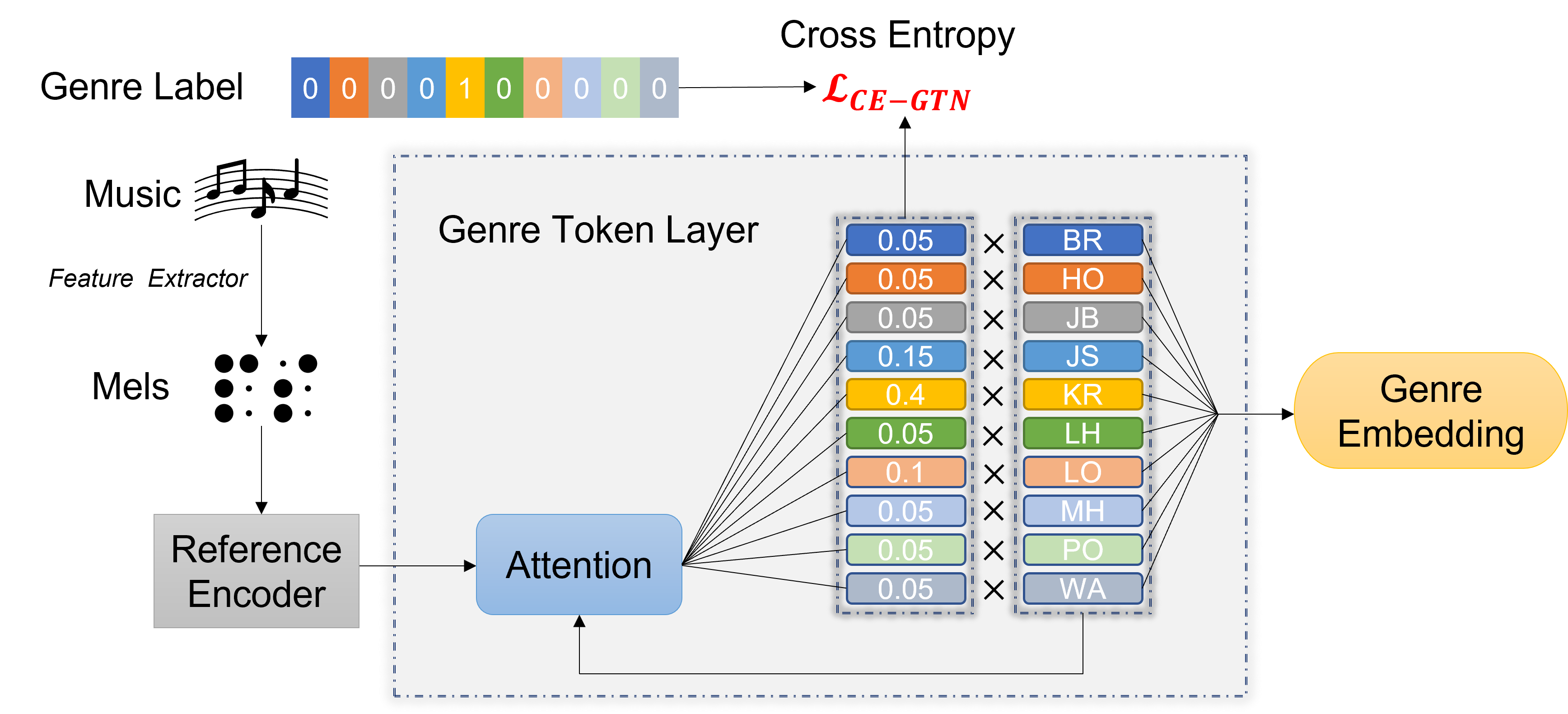}}
\caption{Architecture of the proposed Genre Token Network (GTN).}
\label{fig:pgtn}
\end{figure}

\vspace{-0.1cm}

\section{Proposed Method}
\label{sec:pagestyle}

{\bf Problem Definition.}
We aim to infer genre from music and generate dances based on the inferred genre.
Formally, given a clip of music $X=\left\{x_{t}\right\}_{t=1}^{N}$, 
where ${N}$ is the music length, 
a initial dance pose ${y}_{t=1}$,
our goals are, first, infer the genre $\hat{g} \in G$ from $X$, 
where $G$ is the set of pre-defined genre categories.
Second, use $\hat{g}$ as a condition and $X$ to generate dance $\hat{Y}=\left\{\hat{y}_{t}\right\}_{t=2}^{N}$.

We intend to generate dances with genre consistency. 
In this section, we first describe GTN and show how to infer genre from the music. 
Second, we introduce our genre-consistent dance generation framework.
After that, the strategy of pre-training and fine-tuning for GTN is explained.
The architecture of our proposed GTN-Bailando is shown in Fig.~\ref{fig:framework}.

\vspace{-0.1cm}

\subsection{Genre Token Network}
\label{ssec:subhead}

\vspace{-0.1cm}

Inspired by the work on emotion style tokens based speech synthesis~\cite{YuxuanWang2018StyleTU, WuPengfei2019EndtoEndES}, 
we propose supervised training of genre token to achieve the correlation between genre and music.
As shown in Fig.~\ref{fig:pgtn}, the architecture of the proposed GTN is
composed of three modules: reference encoder, genre token layer, and genre embedding.

The reference encoder is adopted from~\cite{YuxuanWang2018StyleTU}, which is applied to compress the audio signal into a vector of set length.
In this work, the mel-spectrogram of the music clip is fed to the reference encoder, and compressed into the learnable reference embedding. 

The genre token layer includes a set of genre token embeddings and an attention module, which uses the reference embedding as the query vector.
The attention module learns a measure of similarity between the reference embedding and each token in a set of randomly initialized embeddings. 
This set of embeddings, also referred as genre tokens, are shared across all music clips. 

The output of the genre token layer is the probability that the input music belongs to each genre category. 
In order to improve the robustness of the GTN, a soft embedding method~\cite{lei2022msemotts} is used to represent the genre. 
In whis way, the tokens are then weighted summed by the probability to form a genre embedding.



In order to enhance the correlation between music and genre, the number of tokens is set to be consistent with the number of genre categories.
Meanwhile, the genre label is converted to one-hot embedding and introduced into the genre token layer to serve as targets for token weights. 
Thereby, GTN is optimized via supervised training with cross-entropy loss between the genre labels and the genre token weights as follow,

\vspace{-0.2cm}

\begin{equation}
    \mathcal{L}_{C E-GTN}=\frac{1}{T} \sum_{t=0}^{T-1}  \text { CE }\left({g}_{t}, \hat{g}_{t}\right) ,
\end{equation}

\vspace{-0.2cm}

where ${g}_{t}$ and $\hat{g}_{t}$ denote the genre label vector and the genre token weight vector of the music for $t$-th time's clip. 
$\text { CE }(\cdot)$ refers to the cross entropy loss function, $T$ represents the number of music clips.

\vspace{-0.1cm}

\subsection{Dance Generation Framework}
\label{ssec:subhead}

\vspace{-0.1cm}

Inspired by Bailando~\cite{siyao2022bailando}, we generate dances based on VQ-VAE with cross-conditional GPT. 
Since there is a correlation between the velocity of the dance movement and the musical energy, we further consider the energy feature to improve the motion quality of the generated dance.

Precisely, given a piece of music as the input, we first extract the energy and musical features, then embed them to learnable vector $\boldsymbol{Z}_{e}$ and $\boldsymbol{Z}_{m}$, respectively.
In the meantime, we extract the mel-spectrogram of the input music clip, then send it to the GTN to produce the genre embedding $\boldsymbol{Z}_{g}$. 
Then, we concatenate $\boldsymbol{Z}_{m}$ and $\boldsymbol{Z}_{e}$ on the temporal dimension, and add $\boldsymbol{Z}_{g}$, forming $\mathbf{m}$.
As for the dance, we first feed the skeleton joints position of the dance clip into the VQ-VAE encoder to generate the upper and lower half-body pose codes, then embed them to learnable vector $\mathbf{u}$ and $\mathbf{l}$, respectively.

After getting $\mathbf{m}$,$\mathbf{u}$,$\mathbf{l}$, we concatenate them on the temporal dimension and add a learned positional embedding, then feed it to the cross-conditional GPT. 
At last, we get the outputs of the GPT, which are the probabilities of the upper and lower half-body pose codes.
We get the predicted upper and lower body pose code pairs from the probabilities and feed them into the VQ-VAE decoder to get the future dance. 

Note that, we use the teacher-forcing scheme for GTN during training to improve the overall genre consistency of the dance generation framework.
The cross-conditional GPT is optimized via supervised training with cross-entropy loss between the predicted action probability $\mathbf{a}$ and ground-truth pose codes $p$ as follow:

\vspace{-0.2cm}

\begin{equation}
\mathcal{L}_{C E-GPT}=\frac{1}{T^{\prime}} \sum_{t=0}^{T^{\prime}-1} \sum_{h=u, l} \text { CE }\left(\mathbf{a}_{t}^{h}, p_{t+1}^{h}\right).
\end{equation}

\vspace{-0.1cm}

Finally, the loss of dance generation framework $\mathcal{L}_{DG}$ can be computed as:

\vspace{-0.2cm}

\begin{equation}
\label{loss}
    \mathcal{L}_{DG} = \alpha \cdot \mathcal{L}_{C E-GTN} + \beta  \cdot \mathcal{L}_{C E-GPT} .
\end{equation}

\vspace{-0.3cm}


\subsection{Pre-training and Fine-tuning}
Although GTN is capable of inferring the genre, it still suffers from poor generalization capability, 
making it difficult to establish the correlation between the genre and the music nor to ensure genre consistency in generated dances.
It is due to the fact that the music training data in the dance-music-aligned dataset is not large enough (less than 1 hour in length for all the ten genres).

Therefore, we propose a pre-training method to enhance the generalization capability for GTN and improve genre consistency for generated dances.
First, we collect a large amount of dance background music data from the Internet, tag corresponding dance genre labels, and then use this data to train the GTN.
After that, we load the parameters of the pre-trained GTN for fine-tuning while training the cross-conditional GPT of the dance generation framework.
Note that, in order to prevent GTN from overfitting during fine-tuning, the GTN will freeze when it reaches the threshold epoch.



\vspace{+0.1cm}

\section{Experiments}
\label{sec:typestyle}

\subsection{Datasets}
\label{ssec:subhead}

\subsubsection{Dataset for pre-training GTN}

In pre-training GTN, we collect the background music of the dance corresponding to genres from the Internet. 
We collect 1-hour music data for each genre, constituting a music dataset of ten genres of 10 hours in length.
The genres are the same as the ten genres in the dance-music-aligned dataset AIST++~\cite{li2021dance},
which are Break (BR), Ballet Jazz (JB), House (HO), Street Jazz (JS), Krump (KR), LA style Hip-hop (LH), Lock (LO), Middle Hip-hop (MH), Pop (PO), and Waack (WA).

\subsubsection{Dataset for dance generation}

During training and evaluating the dance generation framework with fine-tuning GTN, we use the AIST++~\cite{li2021dance} dataset, which is the largest publicly available 3D dance-music-aligned dataset with genre labels to our knowledge. 
The AIST++ dataset includes 992 high-performance 60-FPS 3D pose sequences in SMPL format, split into 952 and 40 as training and validation sets, respectively.

\subsection{Experiment Setup}
For GTN, we extract an 80-dimensional mel-spectrogram from the music by Librosa.
For the dance generation framework, the musical features are extracted by Librosa, including mel frequency cepstral coefficients (MFCC), MFCC delta, constant-Q chromagram,tempogram, and onset strength, resulting in a 438-dimensional musical feature.
Meanwhile, we extracted energy from the music by referring to Fastspeech2~\cite{ren2020fastspeech}, resulting in a 1-dimensional energy feature.

The settings of dance generation framework and GTN follow previous studies, and we set $\alpha = 1$, $\beta = 0.001$ for equation~(\ref{loss}). 
The num of tokens is set to 10 in the genre token layer in GTN. 
The VQ-VAE in the framework is adopted from the choreographic memories in Bailando~\cite{siyao2022bailando}.
All dance and music data are cropped to the length of 4 seconds.
We first pre-train GTN for 250 epochs and VQ-VAE for 500 epochs, then load the pre-trained VQ-VAE to generate the pose codes corresponding to the dance motions.
After that, we load the pre-trained GTN and train the dance generation framework for 400 epochs, and after 90 epochs, we freeze the GTN.

We compare our proposed method with FACT~\cite{li2021dance} and Bailando~\cite{siyao2022bailando}, which are among the state-of-the-art dance generation methods.
For each method, we generate dances in the condition of ten different AIST++ test music and two different starting pose codes or sequences, which results in 20 dance clips. We cut the generated dances into the length of 20 seconds for further experiments.

\vspace{-0.4cm}

\subsection{Experimental Results and Analysis}
\label{ssec:subhead}

\vspace{-0.1cm}

\subsubsection{Genre Embedding Visualization}
\label{sssec:subsubhead}

\vspace{-0.2cm}

\begin{figure}[htb]
\centering
\centerline{\includegraphics[width=7cm]{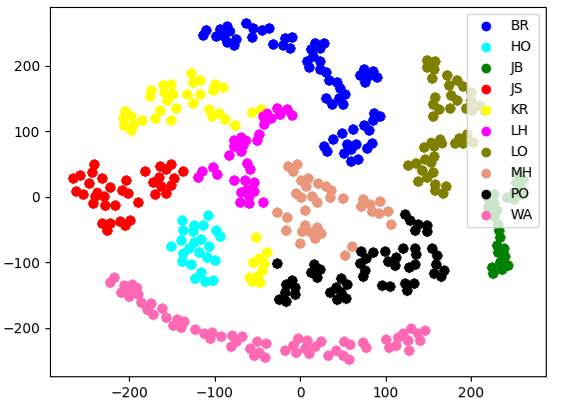}}
\vspace{-0.3cm}
\caption{T-SNE visualization of genre embedding. Different colors mark different dance genres.}
\label{fig:1}
\end{figure}

In Fig.~\ref{fig:1}, we illustrate genre embedding of different genres visualized by t-SNE method.
We use the AIST++ test set to verify the GTN in the dance generation framework, which shows that the different genre embeddings are well separated from each other, proving that the GTN can infer genre from music properly.

\vspace{-0.2cm}

\vspace{-0.6cm}




\begin{table*}[th]
\small
 \centering
 \setlength{\tabcolsep}{8pt}
 \caption{Evaluation results of different dance generation frameworks. 
Dance quality and genre consistency are results of MOS with 95\% confidence intervals. `*' denotes the proposed model.
`w/o' is short for `without' in ablation study.}
 \begin{tabular}{ccccccccc}\toprule
    & \multicolumn{2}{c}{Motion Quality} & \multicolumn{2}{c}{Motion Diversity} & & \multicolumn{2}{c}{User Study}
    \\\cmidrule(lr){2-3}\cmidrule(lr){4-5}\cmidrule(lr){7-8}
             & $FID_{k}\downarrow$  & $FID_{g}\downarrow$ & $DIV_{k}\uparrow$ & $DIV_{g}\uparrow$ & $BAS\uparrow$ & Quality$\uparrow$ & Consistency$\uparrow$\\\midrule
    Ground-Truth                & 17.10 & 10.60 & 8.19 & 7.45 & 0.2484 & -             & -            \\\cmidrule(lr){1-8}
    FACT                        & 37.31 & 34.87 & 5.75 & 5.47 & 0.2175 & 2.15$\pm$0.10 & 2.19$\pm$0.11\\
    Bailando                    & 32.11 &  9.95 & 6.09 & 5.70 & 0.2299 & 3.19$\pm$0.11 & 3.11$\pm$0.11\\
    Proposed*                   & \textbf{29.51} &  \textbf{8.57} & \textbf{6.15} & \textbf{6.65} & \textbf{0.2352} & \textbf{3.65$\pm$0.10} & \textbf{3.60$\pm$0.10}\\\cmidrule(lr){1-8}
    w/o $\mathcal{L}_{C E-GTN}$ & 34.35 & 10.92 & 5.51 & 5.95 & 0.2292 & 3.35$\pm$0.13 & 3.25$\pm$0.13\\
    w/o teacher-forcing         & 41.77 & 14.13 & 5.96 & 3.54 & 0.2298 & 3.49$\pm$0.13 & 3.29$\pm$0.14\\
    w/o pre-trained GTN         & 30.55 &  7.03 & 7.03 & 6.60 & 0.2466 & 3.10$\pm$0.14 & 3.06$\pm$0.15\\
    w/o energy                  & 32.66 &  6.23 & 6.23 & 6.94 & 0.2366 & 3.40$\pm$0.15 & 3.33$\pm$0.15
    \\\bottomrule
 \end{tabular}
 \label{exp}
\end{table*}

\vspace{-0.1cm}

\begin{figure*}[th]
\centering
\centerline{\includegraphics[width=17cm]{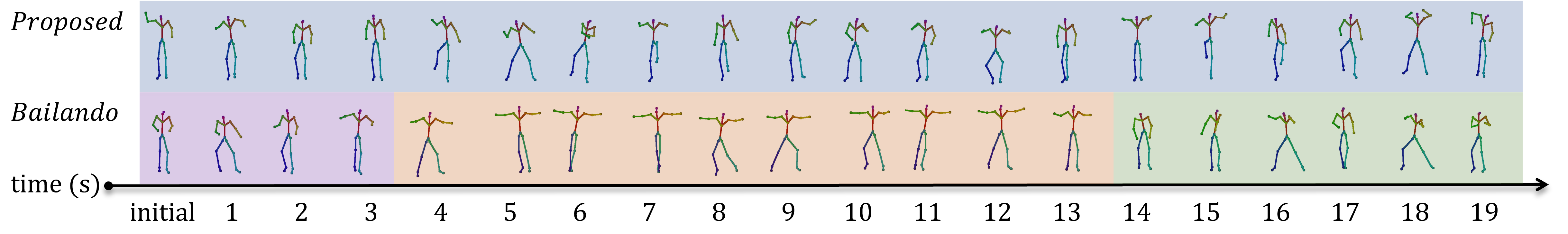}}
\vspace{-0.4cm}
\caption{Genre consistent visualization. Different colors mark dances of different genres.}
\label{fig:control}
\end{figure*}


\subsubsection{Subjective Evaluation}
\label{sssec:subsubhead}

\vspace{-0.1cm}

Due to the lack of objective quantitative metrics for genre consistency,
we conduct subjective evaluation referring to the method in emotional speech synthesis\cite{WuPengfei2019EndtoEndES, lei2022msemotts},
to further evaluate the visual performance of dances generated by the proposed dance generation framework. 
The test is conducted with 24 subjects, who have participated in the dance training and have a certain understanding of the included dance genres. 
The subjects are asked to evaluate the dance quality and genre consistency, and rate the dances on a scale of 1-5 (the poorest score is 1, the best score is 5) with a 1-point interval.

The last two columns of Table~\ref{exp} report the mean opinion scores (MOS) on dance quality and genre consistency.
The proposed model outperforms all baseline models, 
indicating that GTN can establish the correlation between music and genres.
Also, with the inferred genre to serve as a condition, the proposed dance generation framework can generate higher-quality and genre-consistent dances.


\vspace{-0.5cm}

\subsubsection{Objective Evaluation}
\label{sssec:subsubhead}

\vspace{-0.1cm}

For objective evaluation, following~\cite{siyao2022bailando}, we evaluate the quality and the diversity of the generated dance motions, and the alignment of the generated motions to the music beats. 
Specifically, as for the motion quality, we calculate the distribution distance between all the motion sequences of AIST++ and the generated dance motions through Fréchet Inception Distance ($FID$)~\cite{MartinHeusel2017GANsTB} on the kinetic feature (noted as '$FID_{k}$') and the geometric feature (noted as '$FID_{g}$'). 
The lower the $FID$ is, the closer the generated dances are to the ground-truth.
As for the motion diversity, we calculate the average Euclidean distance on the kinetic feature (noted as '$DIV_{k}$') and the geometric feature (noted as '$DIV_{g}$'). 
The higher the $DIV$ is, the larger the distance, meaning that the generated dances have more various movements.
As for the alignment, we calculate the Beat Align Score ($BAS$) between the music beats and the motion beats. 
The higher the $BAS$, the more the dance is on the beat.
The results of different methods as shown in Table~\ref{exp}.

It can be seen that our proposed framework outperforms baseline frameworks on all aspects. 
Specifically, our proposed dance generation framework achieves the lowest $FID_{k}$ and $FID_{g}$, as well as the highest $DIV_{k}$, $DIV_{g}$ and $BAS$. 
The proposed method generates higher-quality, more diverse dances by taking genre and energy into account, as well as improves the alignment between dance movements and the musical melody.

\vspace{-0.2cm}

\subsubsection{Ablation Study}
\label{sssec:subsubhead}

\vspace{-0.1cm}

Moreover, we conduct ablation study to explore the ability of our model to produce dances that are realistic. The results are shown in Table~\ref{exp} and visual comparisons of the ablation study can also be found in the demo webpage.

As shown in Table~\ref{exp}, when the framework is optimized without the $\mathcal{L}_{C E-GTN}$ and the teacher-forcing scheme is not used during the training, the generated dances will be much dissimilar to the ground truth dances and have lower diversity.
When the correlation between the velocity of the dance movement and the musical energy is not considered, the expressiveness of the dance will reduce.
When the pre-training and fine-tuning strategy is not used, although the framework can generate dances that are similar to ground truth, it will have limitations in genre consistency due to the poor generalization capability of GTN.

\vspace{-0.2cm}

\subsubsection{Genre Consistent Visualization}
\label{sssec:subsubhead}

\vspace{-0.1cm}

To further evaluate the genre consistency of the generated dances, we visualize the qualitative results from the joint skeletons of dances generated by our proposed framework and Bailando~\cite{siyao2022bailando}.
We randomly select a 20-second clip from the ``LO"~(Locking) genre dances generated by each framework, and sample the result at a frequency of 1 FPS. 

As shown in Fig.~\ref{fig:control}, given a music clip, the dance generated by Bailando performs multiple dance genres, which is inappropriate with the music.
While, our framework can infer the genre and generate a dance that matches the melody of the music and is consistent with the ``LO" genre. 
More genre-consistent visualization results can be found in the \href{https://im1eon.github.io/ICASSP23-GTNB-DG/}{demo-page}.

\vspace{-0.1cm}

\section{Conclusion}
\label{sec:majhead}


In this paper, to address the challenge of genre consistency, we propose a dance generation framework, GTN-Bailando, which is able to establish the correlation between music and genres, and generate high-quality, genre-consistent dances on the basis of energy, music, and the genre inferred through GTN.
Meanwhile, the proposed strategy of pre-training GTN on the large-scale dataset effectively improves its generalization ability.
Experimental results on AIST++ show that our method can generate high-quality and genre-consistent dances based on the music.

\textbf{Acknowledgment:}
This work is supported by National Natural Science Foundation of China (62076144), 
Shenzhen Key Laboratory of next generation interactive media innovative
technology (ZDSYS 20210623092001004) and Shenzhen Science and Technology
Program  (WDZC20220816140515001, JCYJ20220818101014030).


\clearpage

\bibliographystyle{IEEEbib}
\bibliography{strings,refs}

\end{document}